\documentclass{elsart}

\usepackage{graphicx}
\usepackage{amsmath}
\usepackage{amssymb}

\begin{document}

\begin{frontmatter}

\title{Quasiparticle excitations in frustrated antiferromagnets}

\author[add1]{Adolfo E. Trumper},
\author[add1]{Claudio J. Gazza}, and
\author[add1]{Luis O. Manuel\thanksref{thank1}}
\address[add1]{Instituto de F\'{\i}sica Rosario (CONICET) and
Universidad Nacional de Rosario,
Boulevard 27 de Febrero 210 bis, (2000) Rosario, Argentina\\}

\thanks[thank1]{Corresponding author. Fax: +54-341-4821772. E-mail: manuel@ifir.edu.ar}

\begin{abstract}
We have computed the quasiparticle wave function corresponding
to a  hole injected  in a triangular
antiferromagnet. We have taken into account
multi-magnon contributions within the self consistent Born approximation.
We have found qualitative differences, under sign reversal of
the integral transfer $t$, regarding the multi-magnon
components and the own existence of
the quasiparticle excitations. Such differences are due to the
subtle interplay between magnon-assisted and free hopping
mechanisms. We conclude that the conventional quasiparticle
picture can be broken by geometrical frustration without invoking
spin liquid phases.
\end{abstract}

\end{frontmatter}

\newpage
During a long time the dynamics of a single hole in an antiferromagnet has been
intensively studied\cite{brinkman70}. Such interest was renewed due
to Anderson's proposition\cite{anderson87} about the probable
existence of non-conventional quasiparticle excitations once a Mott
insulating state is doped. For the single hole case, it was
argued\cite{anderson90} that an induced dipolar distortion
on the magnetic background leads to an orthogonality catastrophe, implying
the vanishing of the quasiparticle weight $z_{\bf k}$.
Subsequent studies\cite{reiter94}, based on exact
diagonalization and the self consistent Born approximation (SCBA)
showed that the existence of such a dipolar distortion can be compatible
with a quasiparticle weight $z_{\bf k}\neq 0$ for the whole Brillouin zone.
Angle resolved photoemission spectroscopy experiments
on Mott insulators confirmed this scenario although the difficulty
to distinguish a coherent $\delta$-function peak  from an incoherent part
of the spectra has risen many controversies about the interpretation of
the available data\cite{wells95}.

Actually, the search for non conventional quasiparticles
is a central subject of the resonance valence bond (RVB) scenario\cite{anderson87}
for the unconventional superconductors. The RVB states were believed to be the
true magnetic ground state of a frustrated triangular antiferromagnet.
Extensive studies on the triangular Heisenberg model\cite{bernu92}, however, indicated the
presence of a symmetry broken ground state with a $120^{\circ}$ N\'eel order.
In this article we will study the quasiparticle wave function
corresponding to a single hole injected
in an ordered triangular antiferromagnet (AF). From this
wave function it is possible to compute the contribution
of a different number of magnons in the formation of the
quasiparticle.
Preliminary
results regarding the hole spectral functions have been
published in ref. \cite{trumper04}.

To take into
account the coupling of the hole motion with
the spin fluctuations of the magnetic background we use the $t\!-\!J$ model.
We will give firm evidence that even in this magnetic ordered state the
breakdown  of the conventional quasiparticle excitations,
 produced by the proliferation of multimagnon processes,
 is possible.
In contrast to non-frustrated antiferromagnets there are two mechanisms
for hole motion, one magnon-assisted and the other tight-binding
like,   whose interference might favour quasiparticle (QP)
excitations, or not, depending on the sign of the
integral transfer $t$. This crucial role of the $t$ sign is a manifestation of the particle-hole asymmetry
of the triangular $t-J$ model.

We assume a $120^{\circ}$ N\'eel ordered ground state,
characterized by a magnetic wave vector
${\bf Q}=(\frac{4\pi}{3},0)$, and spin waves as the
magnetic low energy excitations.
Recently, it has been shown
the very good agreement of the linear spin wave theory with
exact diagonalization and quantum Monte Carlo predictions
of the triangular Heisenberg model
\cite{trumper00}.
Using the spinless fermion representation\cite{kane89}  we have obtained the following effective Hamiltonian:
\begin{equation}
\label{HSW}
H  =  \sum_{\bf k}  \epsilon_{\bf k}h^{\dagger}_{\bf{k} } h_{\bf{k} }
 + \sum_{\bf q}  \omega_{\bf q} \alpha^{\dagger}_{\bf q} \alpha_{\bf q}
  - \frac{1}{\sqrt{N}}  \sum_{\bf k, q}
\left[M_{\bf kq}h^{\dagger}_{\bf k}
h_{\bf k-q}\alpha_{\bf q} + h.c. \right]
\end{equation}
with $\epsilon_{\bf k}=-t\gamma_{\bf k}$ and
$\omega_{\bf q}=\frac{3}{2}J\sqrt{(1-3\gamma_{\bf q})(1+6\gamma_{\bf q})},
$
the bare hole and magnon dispersions, respectively.
$M_{\bf kq}=i \sqrt{3} t(\beta_{\bf k}v_{-\bf{q}} -\beta_{\bf k-q}u_{\bf q}) $ is the bare hole-magnon vertex interaction with the geometric factors
$\gamma_{\bf k}=\frac{1}{3}\sum_{\bf e}\cos({\bf k}.{\bf e})$ and $
\beta_{\bf k}=\sum_{\bf e}\sin({\bf k}.{\bf e})$
(${\bf e}$'s are the positive vectors to nearest neighbors), and
$u_{\bf q}$ and $v_{\bf q}$ are the usual Boguliubov coefficients.
Notice that the spin wave calculation
is performed in a local spin quantization axis so as to work
with one kind of magnons ref\cite{chubukov94}.
\noindent In the Hamiltonian (\ref{HSW}) the free hopping hole
term implies a finite probability of the hole to move without
emission or absorption of magnons. This is a direct consequence of the underlying
{\it non-collinear} magnetic structure.
The hole-magnon interaction adds a magnon-assisted mechanism for the hole motion.

The effective Hamiltonian (\ref{HSW}) leads to an analytical
expression for the quasiparticle wave function that takes into
account the contribution of different numbers of
magnons involved in the formation of the quasiparticle (magnetic polaron).
In particular, in the self consistent Born approximation the quasiparticle
wave function (WF) results\cite{reiter94}
$$
|\Phi^{n}_{\bf k}\rangle =  z_{\bf k}\left[ h^{\dagger}_{\bf k}
+ \frac{1}{\sqrt{N}}\sum_{{\bf q}_{1}}g_{{\bf k},{\bf q}_{1}}
h^{\dagger}_{{\bf k}-{\bf q}_{1}} \alpha^{\dagger}_{{\bf q}_1}
+... +\right.$$
$$
\left. +\frac{1}{\sqrt{N^n}} \sum_{{\bf q}_{1},.....,{\bf q}_{n}}
g_{{\bf k},{\bf q}_{1}} g_{{\bf k}_1,{\bf q}_{2}}....
g_{{\bf k}_{n-1},{\bf q}_{n}}
h^{\dagger}_{{\bf k}_n}
\alpha^{\dagger}_{{\bf q}_1}...\alpha^{\dagger}_{{\bf q}_n} \right] |AF\rangle,
$$
where ${\bf k}_i={\bf k}-{\bf q}_1-...-{\bf q}_i$, $|AF\rangle$ is the undoped antiferromagnetic
ground state with a $120^{\circ}$ N\'eel order,
\begin{equation}
g_{{\bf k}_{n},{\bf q}_{n+1}}=
M_{{\bf k}_n,{\bf q}_{n+1}} G_{{\bf k}_{n+1}}
(E_{\bf k}-\omega_{{\bf q}_1}-....-\omega_{{\bf q}_{n+1}}),
\end{equation}

with $ G_{\bf k}(\omega)=(\omega -\epsilon_{\bf k}
-\Sigma_{\bf k}(\omega))^{-1}$
and  the quasiparticle energy is defined
by $E_{\bf k}=\Sigma_{\bf k}(E_{\bf k})$. The self energy  and the Green function are
related as\cite{kane89} $$\Sigma_{\bf k}(\omega )=\frac{1}{N}\sum_{\bf q}
\mid M_{\bf kq}\mid^{2} G_{{\bf k}-{\bf q}}
(\omega-\omega_{\bf q}).
$$
\indent An estimation of the number of magnons necessary to have a
reliable quasiparticle wave function is obtained by requiring the condition norm
$ S^{(n)}_{\bf k}= \langle \Phi^{n}_{\bf k}|\Phi^{n}_{\bf k}\rangle=
 \sum^{n}_{m=0} A^{(m)}_{\bf k}=1.$ The coefficient $A^{(m)}_{\bf k}$ is the $m$-magnon
contribution to the quasiparticle wave function and is defined as
\begin{equation}
A^{(m)}_{\bf k}=\frac{z_{\bf k}}{N^{m}}
\sum_{{\bf q}_{1},.....,{\bf q}_{n}}
g^2_{{\bf k},{\bf q}_{1}} g^2_{{\bf k}_1,{\bf q}_{2}}
......g^2_{{\bf k}_{m-1},{\bf q}_{m}},
\label{coeficients}
\end{equation}
while for the particular case
$m=0$, $A^{(0)}_{\bf k} \equiv z_{\bf k}=
\left(1-
\frac{\partial \Sigma_{\bf k}
(\omega)}{\partial \omega}\right)^{-1}\vert_{E_{\bf k}}$\cite{reiter94}.

For the unfrustrated case, it has been shown the good
agreement of the SCBA with exact diagonalization predictions
as well as with the quasiparticle spectra obtained in
ARPES experiments\cite{wells95}.

\begin{figure}[ht]
\vspace*{0.cm}
\includegraphics*[width=0.90\textwidth]{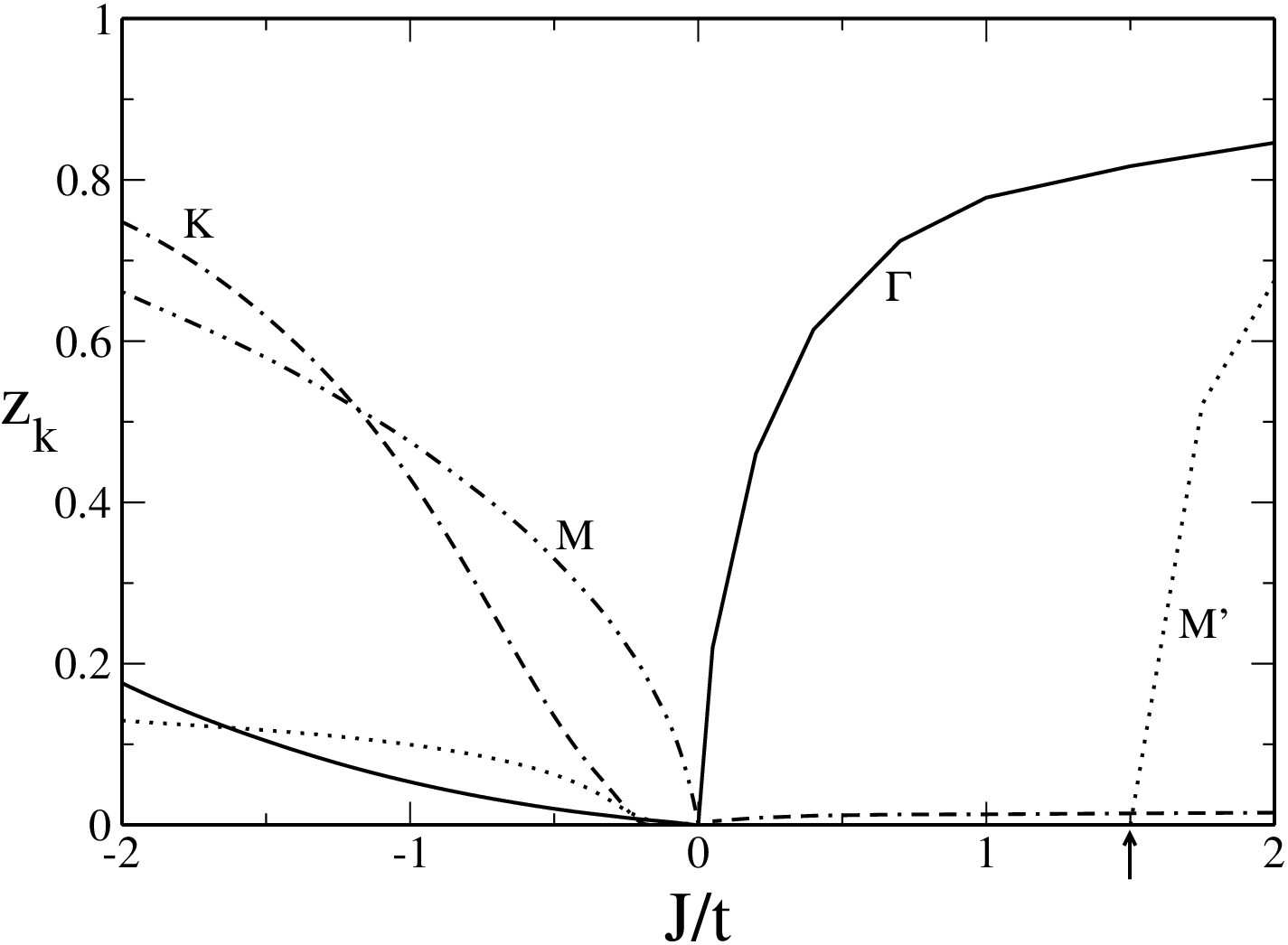}
\caption{}
\label{fig1}
\end{figure}

\begin{figure}[ht]
\vspace*{0.cm}
\includegraphics*[width=0.40\textwidth]{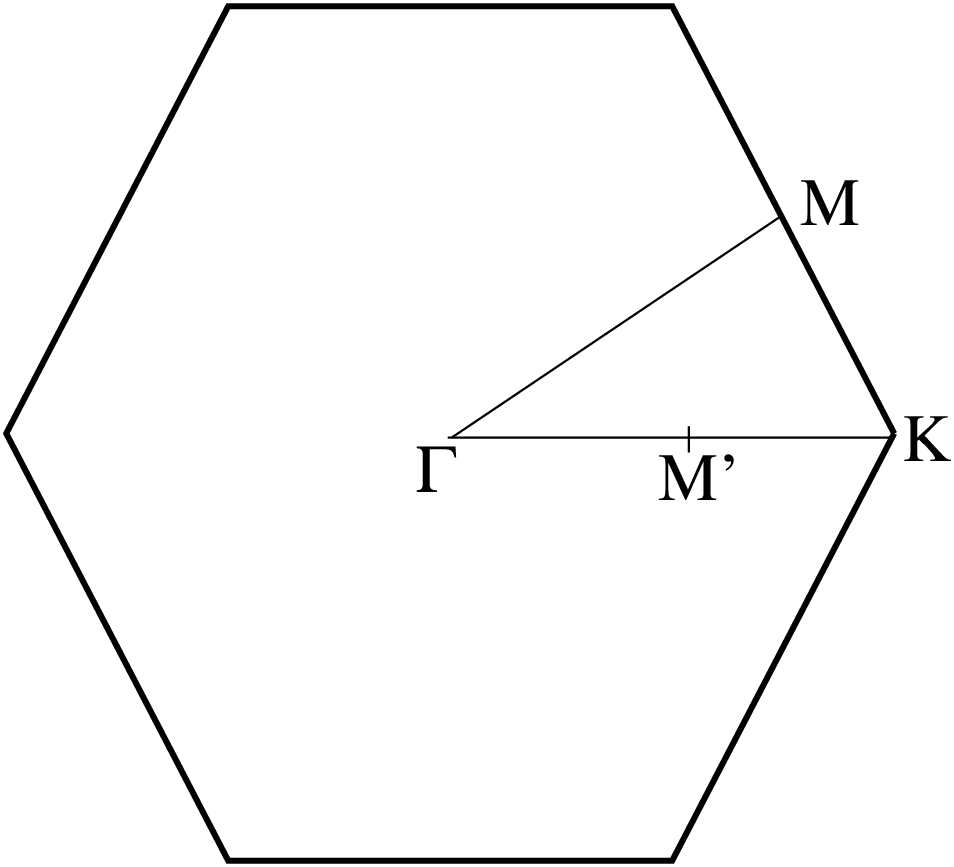}
\caption{}
\label{fig2}
\end{figure}

In a previous work\cite{trumper04} we have checked,
for the frustrated case, the
reliability of the SCBA comparing its results for the hole spectral
functions with exact diagonalization on small size clusters.
We also have extrapolated the quasiparticle weight to the
thermodynamic limit using lattice sizes up to $N=2700$.
 In Fig. \ref{fig1}  we  show the values
of $z_{\bf k}$  as a function of $J/t$ for representative points along high symmetry axes of
the Brillouin zone (Fig. \ref{fig2}).
The most salient feature we have obtained is the vanishing of the
quasiparticle weight
at some momenta for positive $t$. At the $M$ and $M^{\prime}$
points $z_{\bf k}=0$ below $J/t\sim2.5$
and $1.5$, respectively. It is worth noticing the robustness and
then the rapid decay to
zero of $z_{M^{\prime}}$ as $J/t\rightarrow1.5$. A similar behaviour has the QP signal
at $\Gamma$,
the ground state momentum, but it goes to zero only when $J/t \rightarrow 0$. Finally, the QP
signal is very weak, but finite, at the magnetic wave vector $K$ for all the $J/t$ range studied;
for instance, $z_{\bf k}\sim 0.008$ when $J/t=10$.
Surprisingly, for  $J >> t$, $z_{ K}$ does not
seem to approach one as it turns out in the square lattice case\cite{kane89} .
The non existence of quasi-particle excitations is a striking
manifestation of the strong
interference between the free and magnon-assisted  hopping
processes, tuned by the $t$ sign\cite{trumper04}.

On the other hand, for negative $t$  the quasiparticle weight is finite
for all momenta and for $J/t \neq 0$. This behaviour is similar to the one
found in the non-frustrated case \cite{kane89}. It is interesting to
note that now the QP ground state momentum is $M$ ($K$) for
$J/|t| \leq 1.2$ ($J/|t| > 1.2$).
The spectral weight of the QP becomes the most robust for the ground state momenta. The interchange in the ${\bf k}$ dependence of the
 QP weight, evidenced between both signs in Fig. \ref{fig1},
could be thought as a remnant of the particle-hole symmetry that shifts the momenta
by ${\bf Q}$ under $t$ sign reversal.

\begin{figure}[ht]
\vspace*{0.cm}
\includegraphics*[width=0.90\textwidth]{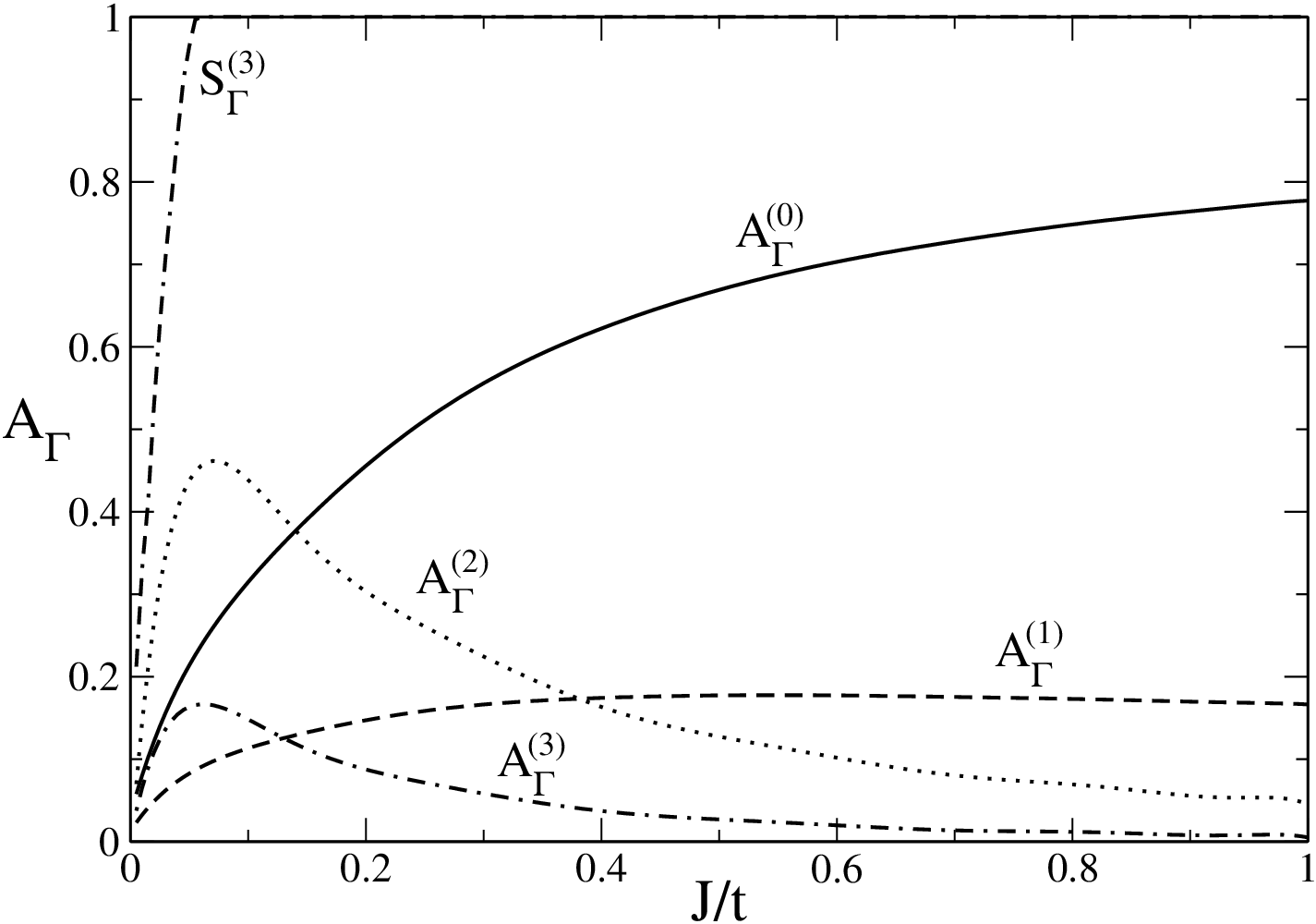}
\caption{}
\label{fig3}
\end{figure}

In what follows we will concentrate on the positive $t$ case.
In order to get some insight of the structure of the quasiparticle
we evaluate the multi-magnon contributions to the QP wave function.
In Fig. \ref{fig3} we show the coefficients $A^{(m)}_{\bf k}$ for
$m=0,1,2,3$ , and their sum $S^{(3)}_{\bf k}$,
for positive $t$ and the ground state momentum, $\Gamma$.
In the weak coupling regime, $J \gg t$, (not shown in the figure)
 the zero magnon
coefficient is the
only relevant contribution to the magnetic polaron WF. As $J/t$ is
lowered the one-magnon and the multi-magnon coefficients start to
 become important. In particular, for $J/t\sim0.1$ $A^{(2)}_{\bf k}$
 is larger than $A^{(0)}_{\bf k}$. It is remarkable that up to
$J/t = 0.05$ the condition norm $S^{(3)}_{\bf k}=1$ is fulfilled, signaling that it is enough to
 consider only the first three magnon terms in the WF.
 As $J/t$ tends to zero all the calculated
coefficients decrease and therefore it would be necessary
to consider other multi-magnon contributions. This proliferation
of magnons close to  $J/t=0$ implies an increasing effective mass
 of the magnetic polaron.

\begin{figure}[ht]
\vspace*{0.cm}
\includegraphics*[width=0.90\textwidth]{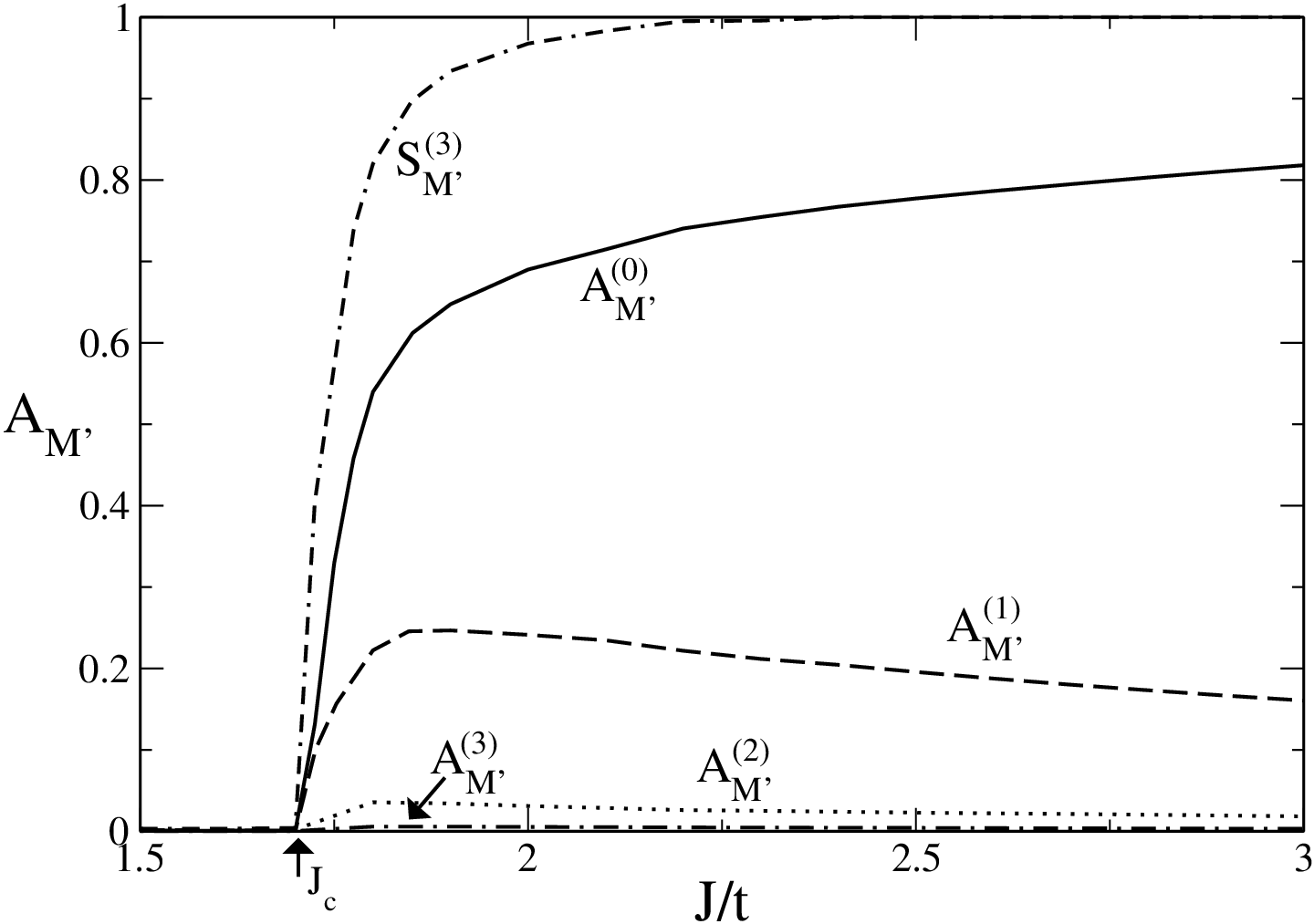}
\caption{}
\label{fig4}
\end{figure}

On the other hand we have made the same analysis for
the $M'$ point of the Brillouin zone, where the quasiparticle
vanishes for a finite value of $J/t$. In Fig. \ref{fig4} we show the
behaviour of the $A$'s  coefficients for this momentum. Now
the zero magnon term remains the most relevant one until
it vanishes. In this case for $J/t \ge 2.2$ the norm condition
$S^{(3)}_{\bf k}=1$ is highly fulfilled with the
zero and the one-magnon contributions. Unlike the $\Gamma$ point
behavior of the WF, the calculated
multi-magnon terms are negligible for all $J/t$. This may
imply a greater proliferation of magnons, even for a finite
value of the magnetic energy scale $J$.
Below $J_c$ the wave function becomes unrenormalizable, that is, the quasiparticle vanishes.

In conclusion, we have studied the quasiparticle wave function of a magnetic polaron in a frustrated antiferromagnet.
We chose the triangular $t\!-\!J$ model which at half filling presents a non-collinear magnetic ground state.
For $t$ positive and ample range of $J/t$, the norm condition $S^{(n)}_{\bf k}=1$ is fulfilled including three or
less magnon processes in the wave function.
We have found a strong momentum dependence of the different terms that build up the quasiparticle.
Our main result is that the wave function becomes unrenormalizable, i.e. the quasiparticle weight is zero,  for
a {\it positive} integral  transfer $t$ at some momenta and at finite $J/t$. The destruction of the quasiparticle signal
is produced by a proliferation of multi-magnon processes, even for finite value of the magnetic energy scale $J$.
This remarkable behaviour is a result of the subtle interference between the two possible mechanism for hole motion.
Such interference is inherent to a hole moving in a non-collinear antiferromagnetic background.

A. E. T. and C. J. G. acknowledge partial support from Fundaci\'on Antorchas.

\newpage

Fig. 1 Quasiparticle weight vs $J/t$. The location of the $\Gamma, M^{\prime}, K, M$ is displayed in Fig.\ref{fig2}

Fig. 2 Brillouin zone of the triangular lattice.

Fig. 3 The norm $S^{(3)}_{\bf k}$ and $A_{\bf k}^{(m)}$ coefficients vs $J/t$, for positive $t$ and
the ground state momentum, ${\bf k}=\Gamma$.

Fig. 4 The norm $S^{(3)}_{\bf k}$ and $A_{\bf k}^{(m)}$ coefficients vs $J/t$, for positive $t$
and ${\bf k}=M'$.

\newpage
\newpage
\newpage

\newpage

\end{document}